\documentclass[prb,twocolumn,superscriptaddress]{revtex4}
\usepackage{amsmath}
\usepackage{amssymb}
\usepackage{graphics}
\usepackage{graphicx}
\usepackage{color}
\usepackage{hyperref}

\begin{document}
\title{Superconductivity proximate to antiferromagnetism in a copper-oxide monolayer grown on Bi$_2$Sr$_2$CaCu$_2$O$_{8+\delta}$}
\author{Shuai Wang}
\affiliation{International Center for Quantum Materials, School of Physics, Peking University, Beijing 100871, China}
\author{Long Zhang}
\affiliation{Kavli Institute for Theoretical Sciences, University of Chinese Academy of Sciences, Beijing 100190, China}
\author{Fa Wang}
\affiliation{International Center for Quantum Materials, School of Physics, Peking University, Beijing 100871, China}
\affiliation{Collaborative Innovation Center of Quantum Matter, Beijing 100871, China}

\begin{abstract}
A nodeless superconducting (SC) gap was reported in a recent scanning tunneling spectroscopy experiment of a copper-oxide monolayer grown on the Bi$_2$Sr$_2$CaCu$_2$O$_{8+\delta}$ (Bi2212) substrate [Y. Zhong {\it et al.}, Sci. Bull. {\bf 61}, 1239 (2016)], which is in stark contrast to the nodal $d$-wave pairing gap in the bulk cuprates. Motivated by this experiment, we first show with first-principles calculations that the tetragonal CuO (T-CuO) monolayer on the Bi2212 substrate is more stable than the commonly postulated CuO$_{2}$ structure. The T-CuO monolayer is composed of two CuO$_2$ layers sharing the same O atoms. The band structure is obtained by first-principles calculations, and its strong electron correlation is treated with the renormalized mean-field theory. We argue that one CuO$_2$ sublattice is hole doped while the other sublattice remains half filled and may have antiferromagnetic (AF) order. The doped Cu sublattice can show $d$-wave SC; however, its proximity to the AF Cu sublattice induces a spin-dependent hopping, which splits the Fermi surface and may lead to a full SC gap. Therefore, the nodeless SC gap observed in the experiment could be accounted for by the $d$-wave SC proximity to an AF order, thus it is extrinsic rather than intrinsic to the CuO$_2$ layers.
\end{abstract}

\maketitle

\section{Introduction}

In the 30-year-long research on high-$T_{c}$ superconducting (SC) cuprates \cite{bednorz1986possible}, one of the most significant achievements has been the well-established $d$-wave SC pairing symmetry. Given the intensive controversy on the nature of the pseudogap phase \cite{timusk1999pseudogap, keimer2015quantum} and even the theoretical starting point \cite{zaanen2006towards}, this consensus may be regarded a touchstone of theoretical proposals. In experiments, $d$-wave pairing symmetry was first revealed by the observation of gapless quasiparticle excitations in nuclear magnetic resonance (NMR), thermodynamic properties, angle-resolved photoemission spectroscopy (ARPES) \cite{damascelli2003angle}, and scanning tunneling microscopy (STM) \cite{fischer2007scanning}, and was later decisively established by phase-sensitive experiments on the pairing symmetry for various families of cuprates \cite{tsuei2000pairing}. Theoretically, $d$-wave pairing symmetry was anticipated by different scenarios at the early stage of research long before a consensus was reached in experiments, including the doped Mott insulator theory \cite{kotliar1988superexchange, zhang1988renormalised} and the antiferromagnetic (AF) fluctuation theory \cite{scalapino1986d}.

This consensus was challenged by a recent experiment \cite{zhong2016nodeless}. Copper oxide monolayers were grown on Bi$_{2}$Sr$_{2}$CaCu$_{2}$O$_{8+\delta}$ (Bi2212) substrates with the molecular beam epitaxy (MBE) technique. Both U- and V-shaped gaps were observed with the scanning tunneling spectroscopy in different spatial regions of the copper-oxide monolayers. The U-shaped gaps with sizes ranging from 16 to 30 meV were attributed to nodeless SC gaps, while the V-shaped gaps from 20 to 50 meV were attributed to a pseudogap. Thereby it was argued that the intrinsic SC pairing in the CuO$_{2}$ layers of cuprates is of $s$-wave form as opposed to the commonly accepted $d$-wave form \cite{zhong2016nodeless}.

The experimentally observed nodeless gap motivated several theoretical works \cite{zhu2016proximity, wang2017nodal, zhu2017two, zhu2017topological}, in which the structure of a monolayer CuO$_2$ on top of a Bi2212 substrate (CuO$_{2}$/Bi2212) was assumed. In Refs.~\onlinecite{zhu2016proximity, wang2017nodal} it was demonstrated that the SC states in monolayer CuO$_2$ induced by the proximity to Bi2212 substrate could be nodeless for certain ranges of the SC pairing order parameters. Furthermore, Zhu {\it et al.} \cite{zhu2017two, zhu2017topological} studied an AF ordered CuO$_2$ layer on top of a $d$-wave SC Bi2212 substrate and proposed that the $d$-wave SC can be fully gapped in the presence of a spin-splitting field induced by AF order (see also Ref. \onlinecite{Lu2014}).

In order to address whether the apparently full SC gap is or is not intrinsic, one must first find out whether the chemical composition of the MBE-grown copper-oxide monolayer is the same as the CuO$_{2}$ layers in the bulk cuprates. In Ref.~\onlinecite{zhong2016nodeless} this was assumed to be true, thus implying that the U-shaped gap is intrinsic to the bulk cuprates as well. However, because the chemical composition cannot be fully controlled during MBE growth, the possibility of a tetragonal CuO (T-CuO) monolayer with the same lattice constant cannot be ruled out, which was also pointed out in Ref.~\onlinecite{zhong2016nodeless}.

In this work, we first study the chemical composition of a copper-oxide monolayer on a Bi2212 substrate with first-principles calculations. We show that the formation energy of a T-CuO monolayer grown on a Bi2212 substrate (T-CuO/Bi2212) is significantly lower than that of a CuO$_{2}$/Bi2212 structure, therefore the T-CuO/Bi2212 structure is more stable than the CuO$_2$/Bi2212 structure postulated in Ref.~\onlinecite{zhong2016nodeless}.

The T-CuO monolayer is composed of two sets of CuO$_2$ sublattices (sharing the same O atoms) with Cu residing at different heights, which is shown in the upper panel of Fig.~\ref{fig:structure}. The effective two-band tight-binding model involving the Cu $3d_{x^{2}-y^{2}}$ orbitals and the magnetic exchange interactions are derived from first-principles calculations. The on-site energy difference of the two Cu sublattices is about 80 meV. We show that it is energetically more favorable if all doped holes are injected into one of the Cu sublattices, while the other Cu sublattice remains half filled because of their on-site energy difference and strong on-site Coulomb repulsion. Using renormalized mean field theory (RMFT) of doped Mott insulators \cite{zhang1988renormalised}, we show that $d$-wave superconductivity is formed in the doped Cu sublattice at low temperatures. Moreover, if the local moments in the half-filled sublattice form long-range AF correlation, a spin-dependent hopping term is induced in the doped Cu sublattice and the Fermi surface is split. Depending on the doping concentration and the spin-dependent hopping strength, the split Fermi surfaces may or may not intersect with the nodal lines of the $d$-wave SC gap function, thus the doped Cu sublattice can show either nodal or full SC gaps. Therefore, we conclude that the observed U-shaped SC gap can be accounted for by its proximity to the AF order, and thus it is extrinsic rather than intrinsic to the CuO$_{2}$ layers.

\section{Chemical composition and electronic structure}

First-principles calculations are performed using the density functional theory (DFT) with the projector augmented wave (PAW) method \cite{blochl1994projector, kresse1999from} as implemented in the Vienna \emph{ab initio} simulation package (VASP) \cite{kresse1996efficiency, kresse1996efficient}. The generalized gradient approximation (GGA) proposed by Perdew, Burke, and Ernzerhof (PBE) \cite{perdew1996generalized} for electron exchange-correlation functional is adopted. The kinetic energy cutoff of the plane waves is set to be 400 eV. We employ the experimental lattice constants of Bi2212 ($a=5.414$ \AA, $b=5.418$ \AA, $c=30.89$ {\AA}) \cite{sunshine1988structure} in the bulk, and a copper-oxide monolayer of the T-CuO or CuO$_2$ structure and an about 10-{\AA}-thick vacuum layer on top of a two-unit-cell-thick stoichiometric Bi$_{2}$Sr$_{2}$CaCu$_{2}$O$_{8}$. The atomic positions in the copper-oxide monolayer are optimized by the quasi-Newton algorithm until the force on each atom is smaller than 0.01 eV/\AA. The reciprocal space is sampled using a $9 \times 9 \times 1$ and a $15 \times 15 \times 1$ $\Gamma$-centered Monkhorst-Pack grid \cite{monkhorst1976special} in the structural optimization and the self-consistent static calculations, respectively.

\begin{figure}[!tb]
\centering
\includegraphics[width=0.48\textwidth]{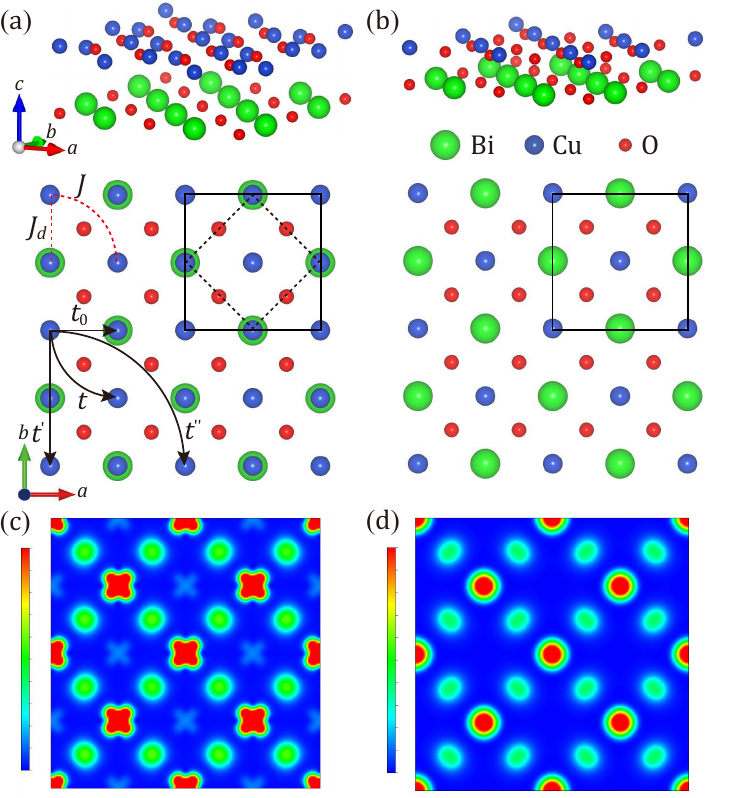}
\caption{Crystal structures and local density of states of T-CuO/Bi2212 and CuO$_2$/Bi2212. Perspective views and top views of (a) T-CuO/Bi2212 and (b) CuO$_2$/Bi2212. Only the topmost BiO layers in the Bi2212 substrates are shown. The supercells used in the calculations (black solid lines) and the primitive cell in the T-CuO monolayer (black dashed lines) are also shown. The $t$-$J$ model parameters are schematically illustrated in (a). (c), (d) Two-dimensional distributions of the integrated local density of states cut at the topmost Cu atom planes of (c) T-CuO monolayer and (d) CuO$_2$ monolayer. The integration window is [$-$0.1, 0.1] eV around the Fermi level.}
\label{fig:structure}
\end{figure}

The relaxed structures of the copper-oxide layers are illustrated in Figs.~\ref{fig:structure}(a) and ~\ref{fig:structure}(b). In the T-CuO monolayer there is an extra set of Cu atoms, which locate over the Bi atoms and are 0.54 {\AA} lower than the Cu atoms locating over the O atoms. We adopt the formation energy to estimate the stability of T-CuO/Bi2212 and CuO$_2$/Bi2212. In the supercell marked by the solid lines in Fig.~\ref{fig:structure}(a) there are two more Cu atoms in the T-CuO/Bi2212 structure than in the CuO$_{2}$/Bi2212 structure. Therefore, the formation energy difference per supercell is given by
\begin{equation} \label{binding}
\Delta E=E_{\text{T-CuO/Bi2212}}-E_{\text{CuO}_2\text{/Bi2212}}-2E_{\text{Cu}},
\end{equation}
in which $E_\mathrm{X}$ represents the total energy of X per supercell estimated by first-principles calculations. The total energies of T-CuO/Bi2212, CuO$_2$/Bi2212, and the face-centered-cubic elemental Cu are $-370.384$, $-361.378$ and $-3.749$ eV, respectively, so the formation energy difference $\Delta E = -1.558$ eV, indicating that the T-CuO/Bi2212 structure is much more stable than CuO$_2$/Bi2212. Therefore we assume that the T-CuO/Bi2212 structure is grown in the experiment and will only study the T-CuO/Bi2212 structure in the remainder of this work.

Figs.~\ref{fig:structure}(c) and ~\ref{fig:structure}(d) show the integrated local density of states (LDOS) cut at the topmost Cu atom plane for each structure, where the integration range of energy is $[-0.1~\mathrm{eV}, 0.1~\mathrm{eV}]$ around the Fermi energy. Due to the 0.54 {\AA} height difference of the two Cu sublattices in the T-CuO monolayer, the lower Cu atoms might be difficult to resolve in STM topography, so the topography of the T-CuO/Bi2212 can be similar to that of CuO$_{2}$/Bi2212. The Cu bilayer structure in the T-CuO surface may in principle be detected by reflection high-energy electron diffraction (RHEED).

\begin{figure*}[!tb]
\centering
\includegraphics[width=0.96\textwidth]{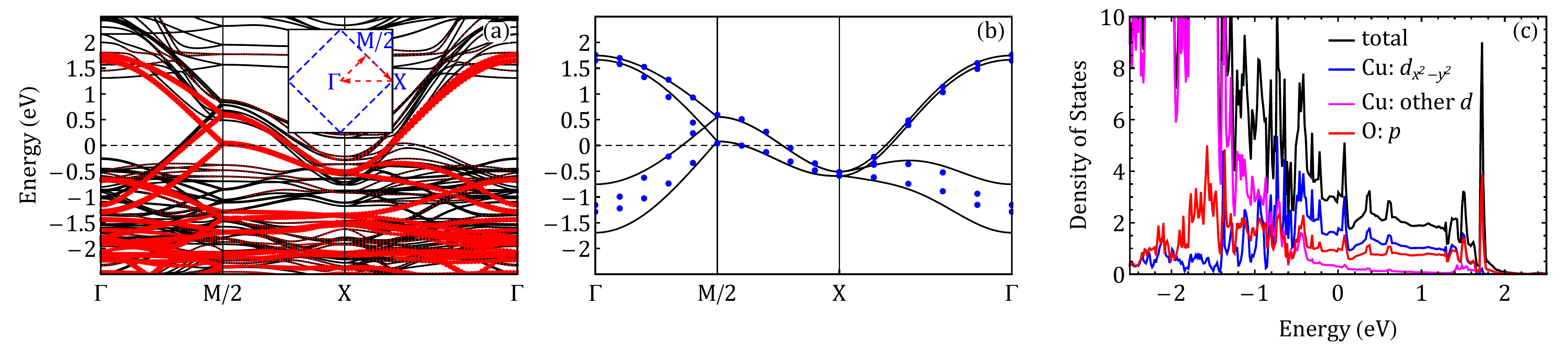}
\caption{Electronic structure and orbital-resolved density of states of the T-CuO/Bi2212 structure. (a) First-principles electronic band structure of T-CuO/Bi2212. The spectral weight contributed by the T-CuO monolayer to each band is indicated by the sizes of the red circles superposed on the energy dispersion. The inset shows the primitive Brillouin zone (black solid rectangle) and the folded Brillouin zone (blue dashed rectangle). (b) Energy dispersion of the T-CuO monolayer from the tight-binding model (black solid lines) is compared with the DFT electronic structure (blue dots). (c) Orbital-resolved density of states in the T-CuO monolayer. The Fermi level is set to $E_\mathrm{F} = 0~\mathrm{eV}$.}
\label{fig:bands}
\end{figure*}

The paramagnetic band structure of T-CuO/Bi2212 in the supercell is shown in Fig.~\ref{fig:bands}(a). The spectral weight contributed by the T-CuO layer to each band is indicated by the sizes of the red circles, from which it is clear that the T-CuO monolayer forms four bands near the Fermi energy, i.e., one band per Cu atom. From the orbital-resolved density of states (DOS) of the T-CuO monolayer shown in Fig.~\ref{fig:bands}(c), the bands near the Fermi energy are predominantly contributed by the Cu $3d_{x^2-y^2}$ orbitals hybridized with the O $2p$ orbitals, which is consistent with the Zhang-Rice singlet (ZRS) \cite{zhang1988effective} formation in the copper-oxide layer. We remark that the T-CuO monolayer can be treated as two CuO$_{2}$ layers sharing the same O atoms, and the $2p_{x}$ and $2p_{y}$ orbitals of one O atom mainly hybridize with the $3d_{x^{2}-y^{2}}$ orbitals of its adjacent Cu atoms along the horizontal and the vertical directions, respectively. As shown in Fig.~\ref{fig:bands}(b), an effective tight-binding model incorporating all the Cu $3d_{x^{2}-y^{2}}$ orbitals can capture the band structure of the T-CuO monolayer around the Fermi energy. The hopping part $H_t$ and the on-site energy part $H_{\mu}$ of this model are
\begin{align} \label{Ht}
\begin{split}
H_t=&-t_0\sum_{[ij],\sigma}d^\dagger_{i,\sigma}d_{j,\sigma}-t\sum_{\langle ij\rangle,\sigma}d^\dagger_{i,\sigma}d_{j,\sigma}\\
&-t^\prime\sum_{\langle ij\rangle',\sigma}d^\dagger_{i,\sigma}d_{j,\sigma}-t^{\prime\prime}\sum_{\langle ij\rangle'',\sigma}d^\dagger_{i,\sigma}d_{j,\sigma},
\end{split}\\
H_{\mu}=&\sum_{i,\sigma}\epsilon_i d_{i,\sigma}^{\dagger}d_{i,\sigma},
\end{align}
where $d^\dagger_{i\sigma}$ creates an electron with spin $\sigma$ at site $i$. The $t_{0}$ term denotes the nearest-neighbor hopping between the two inequivalent Cu sublattices, and the $t$, $t^\prime$, $t^{\prime\prime}$ terms are the hopping between the first, second, and third nearest neighbors within one Cu sublattice. These hopping processes are schematically shown in Fig.~\ref{fig:structure}(a). $\epsilon_i$ is the on-site energy at site $i$, which is different for the two inequivalent Cu sublattices.

Fitting the tight-binding model to the DFT band structure gives the following parameters, $t= 0.366$ eV, $t^\prime=-0.099$ eV, $t^{\prime\prime}=0.059$ eV, and $t_0=-0.117$ eV. The on-site energies at the Cu sites over the O atoms and the Cu sites over the Bi atoms are 0.122 and 0.039 eV, respectively, therefore their on-site energy difference is 83 meV. The energy dispersion from the tight-binding model along the high-symmetry lines is plotted in Fig.~\ref{fig:bands}(b). The hopping parameters are in close agreement with those in the bulk cuprates \cite{andersen1995lda}, and the intersublattice hopping $t_0$ is consistent with that of the T-CuO film grown on the SrTiO$_3$ substrate \cite{moser2014angle, adolphs2016non}.

The effect of the strong on-site Coulomb repulsion of the Cu $3d$ orbitals is twofold. First, double occupation on the same site must be avoided in the hopping process due to the large energy penalty (the single-occupancy constraint), thereby reducing the effective hopping amplitude and the bandwidth, which will be taken care of by the renormalized mean-field theory in the next section. Second, the singly occupied sites form local magnetic moments. The virtual hopping process induces magnetic interactions among the local moments. The magnetic interaction can be derived from the total energies of various prescribed magnetically ordered states in the DFT$+U$ calculations. The details are presented in the Appendix. The results can be captured by the Heisenberg interaction,
\begin{equation} \label{Hs}
H_s=J_d\sum_{[ij]}\vec{S}_i \cdot \vec{S}_j+J\sum_{\langle ij\rangle}\vec{S}_i \cdot \vec{S}_j,
\end{equation}
where the $J_d$ and the $J$ terms are interactions between the nearest-neighbor sites on the same sublattice and on different sublattices, respectively, which are illustrated in Fig.~\ref{fig:structure}(a). The fitted values from the DFT$+U$ calculations are $J_d=-5.6$ meV and $J=119.6$ meV, which are close to those of the T-CuO film on the SrTiO$_3$ substrates \cite{moser2014angle, adolphs2016non}.

\section{Renormalized Mean-Field Theory}

On-site repulsion strongly renormalizes the electronic structure. In a conventional Fermi liquid, the repulsive interaction generally enhances the electron effective mass. With the single-occupancy constraint, the hopping amplitude is effectively reduced by a factor proportional to the concentration of the doped holes. This effect is captured by the renormalized mean-field theory (RMFT) proposed in Ref. \onlinecite{zhang1988renormalised}. The RMFT is based on the Gutzwiller approximation \cite{gutzwiller1963effect} of the projected Bardeen-Cooper-Schrieffer (BCS) mean-field wave function \cite{anderson1987resonating}, $|\Psi_{\mathrm{G}}\rangle =P_{\mathrm{G}}|\Psi_{0}\rangle$, in which $|\Psi_{0}\rangle$ is the BCS wave function, and $P_{\mathrm{G}}=\prod_{i}(1-n_{i\uparrow}n_{i\downarrow})$ is the Gutzwiller projection operator that enforces the single-occupancy constraint. The expectation values of the hopping and the magnetic exchange terms in the Hamiltonian are approximated by \cite{gutzwiller1963effect, zhang1988renormalised}
\begin{align}
\langle \Psi_{\mathrm{G}}|H_{t}|\Psi_{\mathrm{G}}\rangle &=g_{t}\langle \Psi_{0}|H_{t}|\Psi_{0}\rangle,\\
\langle \Psi_{\mathrm{G}}|H_{s}|\Psi_{\mathrm{G}}\rangle &=g_{s}\langle \Psi_{0}|H_{s}|\Psi_{0}\rangle.
\end{align}
In other words, the single-occupancy constraint in the $t$-$J$ model $H_{tJ}=H_{t}+H_{\mu}+H_{s}$ is relaxed, and the price to pay is to replace the original Hamiltonian with
\begin{equation}
H_{\mathrm{R}}=g_{t}H_{t}+H_{\mu}+g_{s}H_{s}.
\end{equation}
The renormalization factors are given by \cite{zhang1988renormalised}
\begin{equation} \label{gtgs}
g_t=\frac{2\delta}{1+\delta}, \quad g_s=\frac{4}{(1+\delta)^2},
\end{equation}
where $\delta$ is the hole concentration away from half filling. Approaching half filling, the effective bandwidth is progressively reduced by the renormalization factor $g_{t}$, which captures the Brinkman-Rice scenario of a diverging effective mass at the Mott transition \cite{brinkmann1970application}.

In the T-CuO monolayer, the doping concentrations on the two Cu sublattices are different due to the 83-meV on-site energy difference. Moreover, for a small doping concentration, the doped holes tend to be injected into one of the Cu sublattices, leaving the other Cu sublattice undoped. This assumption can be justified with the RMFT of the $t$-$J$ Hamiltonian. If both Cu sublattices are treated on an equal footing, one sublattice would be hole doped while the other sublattice would be slightly electron doped. For example, doping 0.1 holes per Cu into the T-CuO monolayer results in 0.22 holes per Cu in the Cu sublattice over the O atoms and 0.02 electrons per Cu in the Cu sublattice over the Bi atoms. The strong on-site repulsion would push the excess electrons away, leaving one of the Cu sublattices half filled. This is reminiscent of the orbital-selective Mott transition scenario in multiband transition-metal compounds \cite{anisimov2002orbital}. Therefore, we treat the doped T-CuO monolayer as a doped Cu square lattice proximate to an undoped Cu lattice, the latter of which may have AF long-range order.

\begin{figure}[!tb]
\centering
\includegraphics[width=0.4\textwidth]{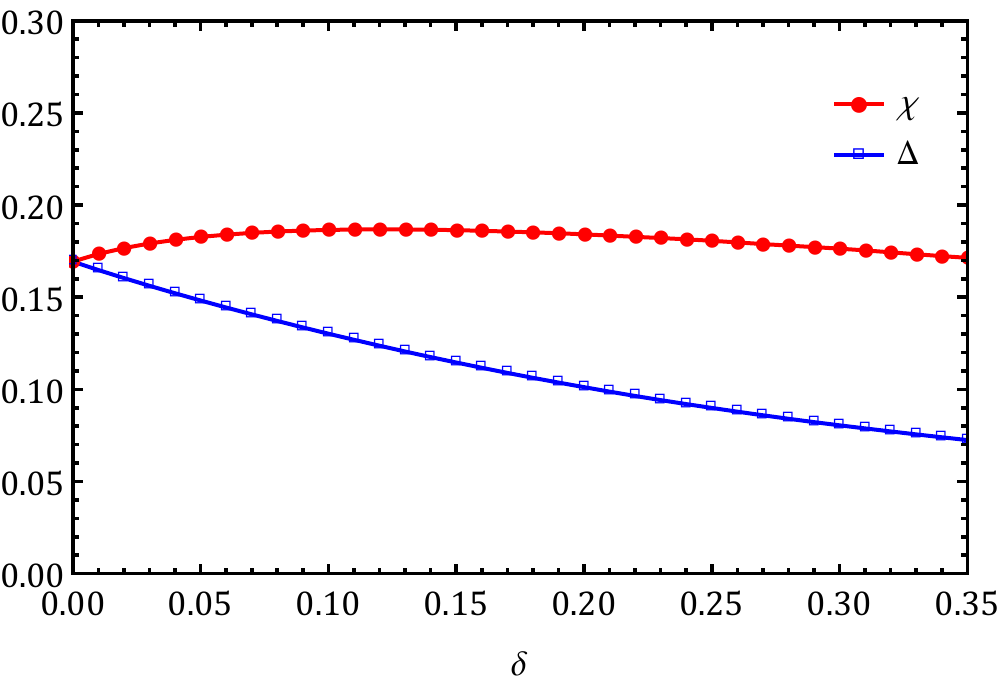}
\caption{Doping dependence of the mean field parameters $\chi=\langle d^\dagger_{i\sigma}d_{i+x\sigma}\rangle_{0}$ and $\Delta=\langle \sigma d_{i+x-\sigma}d_{i\sigma}\rangle_{0}$. The hole doping concentration $\delta$ is related to the band filling by $n=1-\delta$.}
\label{fig:deltachi}
\end{figure}

The RMFT Hamiltonian for the doped Cu sublattice is \cite{zhang1988renormalised, j?drak2010consistent}
\begin{equation} \label{Hmf}
\begin{split}
H_{\mathrm{R}}'=&-g_tt\sum_{\langle ij\rangle,\sigma}d^\dagger_{i\sigma}d_{j\sigma}-g_{t}t^\prime\sum_{\langle ij\rangle^\prime,\sigma}d^\dagger_{i\sigma}d_{j\sigma}\\
&-g_{t}t^{\prime\prime}\sum_{\langle ij\rangle^{\prime\prime},\sigma}d^\dagger_{i\sigma}d_{j\sigma}-\mu\sum_{i,\sigma} d^{\dagger}_{i\sigma}d_{i\sigma}\\
&-\frac{3}{4}g_s J \sum_{\langle ij\rangle}\big(\chi_{ji}d^\dagger_{i\sigma}d_{j\sigma}+\Delta_{ji}d^\dagger_{i\sigma}d^\dagger_{j-\sigma}+\mathrm{h.c.}\big),
\end{split}
\end{equation}
in which $\Delta_{ij}=\langle \sigma d_{j-\sigma}d_{i\sigma}\rangle_{0}$, $\chi_{ij}=\langle d^\dagger_{i\sigma}d_{j\sigma}\rangle_{0}$ are the expectation values over the BCS wave function $|\Psi_{0}\rangle$, and $\mu$ is chemical potential for adjusting the doping concentration. The single-particle excitation gap is determined by the mean-field pairing $\Delta_{ij}$, but the SC order parameter is given by the expectation value on the Gutzwiller projected wave function, $\langle \sigma d_{j-\sigma}d_{i\sigma}\rangle_{\mathrm{G}}=g_{t}\Delta_{ij}$. Therefore, the RMFT approximation implies the separation of two energy scales, the pseudogap and the SC gap \cite{zhang1988renormalised}. $\Delta_{ij}$, $\chi_{ij}$, and $\mu$ are determined by self-consistently solving the Bogoliubov de Gennes (BdG) equations. The SC pairing symmetry is determined by the relative phase of the pairing order parameters along different bond directions. For example, the extended $s$-wave paring is given by $\Delta_{i,i+\hat{x}}=\Delta_{i,i+\hat{y}}=\Delta$, and the $d$-wave pairing is given by $\Delta_{i,i+\hat{x}}=-\Delta_{i,i+\hat{y}}=\Delta$. $\Delta_{ij}$ vanishes for the $s$-wave pairing ansatz in the RMFT calculation. Therefore we expect that the SC state has $d$-wave symmetry. The self-consistent solutions of $\chi$ and $\Delta$ are shown in Fig.~\ref{fig:deltachi}. The single-particle pseudogap $\Delta$ decreases with increasing doping concentration. The SC order parameter in the Gutzwiller approximation is $\Delta_{\mathrm{SC}}=g_{t}\Delta$. $\Delta_{\mathrm{SC}}$ vanishes linearly as the renormalization factor $g_{t}$ approaching the undoped limit. These results are similar to the RMFT of the $t$-$J$ model with the nearest-neighbor hopping term \cite{zhang1988renormalised}.

\section{Fully gapped superconductivity}

\begin{figure}[!tb]
\centering
\includegraphics[width=0.48\textwidth]{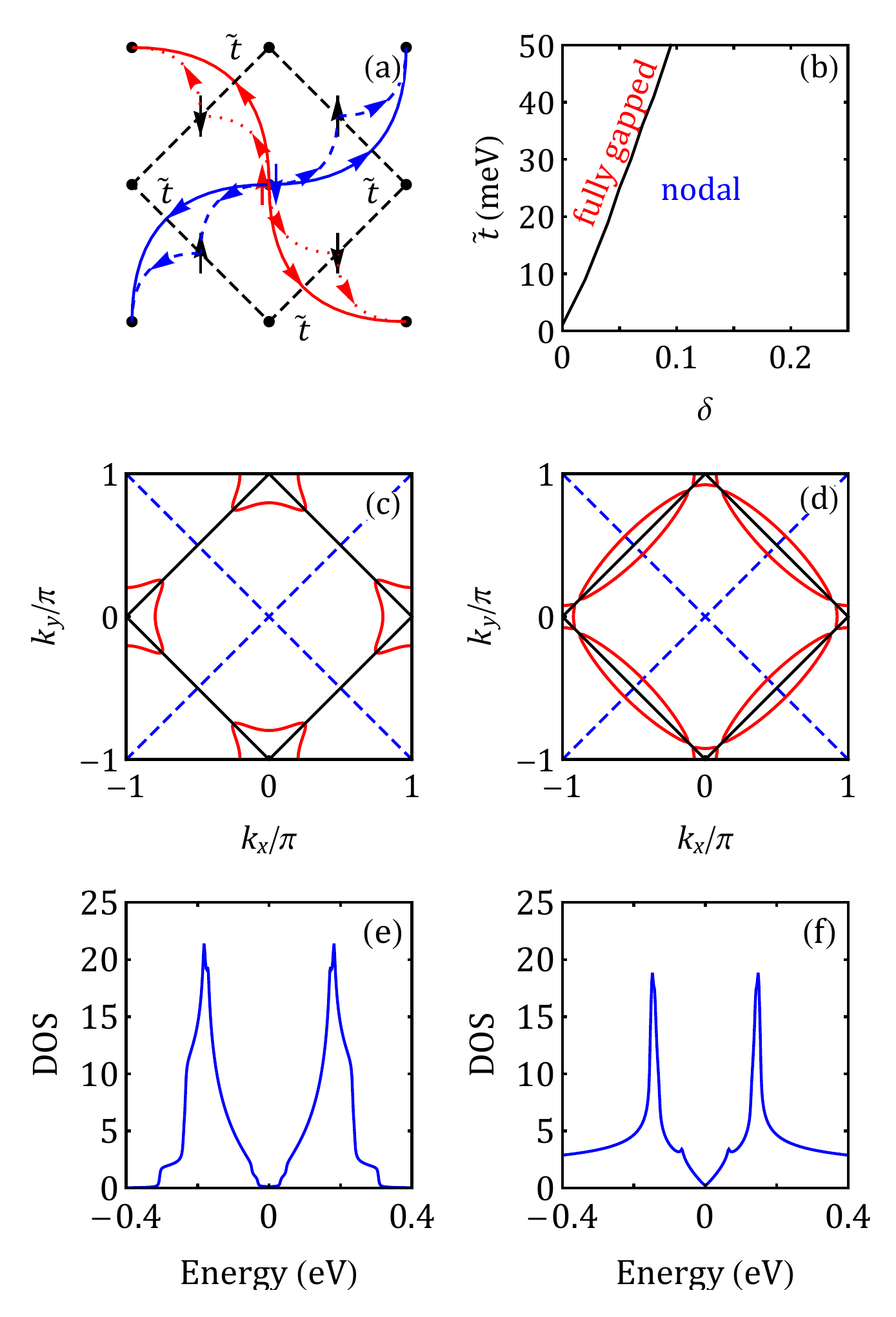}
\caption{Spin-dependent hopping and SC phase diagram. (a) Schematic illustration of the spin-dependent hopping processes in the doped Cu sublattice by its proximity to the AF ordered sublattice. The dashed square encloses the magnetic unit cell with four Cu atoms in the T-CuO monolayer. The sites with and without arrows indicate the AF ordered Cu sublattice and the doped SC Cu sublattice, respectively. (b) Phase diagram in the parameter plane of the doping concentration $\delta$ and the spin-dependent hopping $\tilde{t}$. (c), (d) Fermi surfaces (red solid lines) and nodal lines of the SC gap functions (blue dashed lines) for doping concentrations (c) $\delta=0.01$  and (d) $\delta=0.1$. $\tilde{t}$ is taken to be $20~\mathrm{meV}$. (e), (f) The quasiparticle density of states corresponding to (c) and (d) showing a full SC gap and a nodal V-shaped gap, respectively. The Fermi level is set to $E_\mathrm{F} = 0~\mathrm{eV}$.}
\label{fig:result}
\end{figure}

We then consider the proximity of the doped superconducting Cu sublattice to the undoped Cu sublattice. The proximity induces extra hopping terms in the doped layer via the virtual hopping through the undoped layer, which slightly renormalizes the band structure. However, if the undoped layer has long-range AF order, the possible virtual hopping processes are spin dependent, i.e., the virtual hopping of an electron through a site with the same spin polarization is not allowed due to the Pauli principle, which is illustrated in Fig.~\ref{fig:result}(a). The AF order-induced spin-dependent hopping term doubles the unit cell of the doped layer and breaks the original translational symmetry. Thereby the Cooper pair expectation values would also have such a reduced translational symmetry, and would be modulated in real space consistent with a unit cell doubling.

For simplicity, we assume that the ordered AF moments are along the $z$ direction, and the induced extra hopping terms are fully spin polarized, then this extra spin-dependent hopping term is
\begin{equation}\label{spin}
H_{\tilde{t}}=-\tilde{t}\sum_{\langle ij\rangle',\sigma}\frac{1}{2}(1-\tau^z_{ij}\sigma) d_{i\sigma}^{\dagger}d_{j\sigma}+\mathrm{h.c.},
\end{equation}
in which $\tau^z_{ij}=\pm 1$ is the spin polarization of the intermediate site located in the undoped sublattice between the sites $i$ and $j$ in the doped sublattice.
In terms of the Nambu spinors in the folded Brillouin zone, $\Psi_{\vec{k}}=\big(d_{1\vec{k}\uparrow}, d^\dagger_{1-\vec{k}\downarrow}, d_{2\vec{k}\uparrow}, d^\dagger_{2-\vec{k}\downarrow}\big)^{\mathrm{T}}$, the total Hamiltonian including the RMFT $t$-$J$ Hamiltonian and the spin-dependent hopping term is given by
\begin{equation} \label{Hmf1}
H=\sum_{\vec{k}}\Psi_{\vec{k}}^\dagger
\begin{pmatrix}
\epsilon_{t'}+\tilde{\epsilon}_x	& 0		& \epsilon_t	& \Delta_{\vec{k}}		\\
0	& -\epsilon_{t'}-\tilde{\epsilon}_y 	& \Delta_{\vec{k}}		& -\epsilon_t	\\
\epsilon_t 	& \Delta_{\vec{k}}	& \epsilon_{t'}+\tilde{\epsilon}_y	& 0				\\
\Delta_{\vec{k}}	& -\epsilon_t & 0 & -\epsilon_{t'}-\tilde{\epsilon}_x
\end{pmatrix}\Psi_{\vec{k}},
\end{equation}
in which
\begin{equation} \label{H0}
\begin{split}
&\epsilon_t=-(4g_tt +3 g_s J_2\chi) \cos\frac{k_x}{2} \cos\frac{k_y}{2},\\
&\epsilon_{t'}=-2g_{t}t^\prime(\cos k_x+\cos k_y)-4g_{t} t^{\prime\prime} \cos k_x \cos k_y-\mu,\\
&\tilde{\epsilon}_x=-2\tilde{t}\cos k_x,\quad \tilde{\epsilon}_y=-2\tilde{t}\cos k_y,\\
&\Delta_{\vec{k}}=3g_sJ_2 \Delta \sin\frac{k_{x}}{2}\sin\frac{k_{y}}{2}.
\end{split}
\end{equation}
Diagonalizing the above Hamitonian yields the following quasiparticle excitation spectrum,
\begin{equation} \label{Ek}
E^\pm_{\vec{k}}=\sqrt{\Big(\epsilon_{t'}+\frac{1}{2}(\tilde{\epsilon}_x+\tilde{\epsilon}_y)\pm\frac{1}{2}\sqrt{(\tilde{\epsilon}_x-\tilde{\epsilon}_y)^2+4\epsilon_t^2}\Big)^2+\Delta_{\vec{k}}^{2}}.
\end{equation}
The underlying Fermi surface is folded due to the spin-dependent hopping term, the original Fermi surface is split into two, and the SC gap is formed on the split Fermi surfaces. The Fermi surfaces and the nodal lines of the gap function $\Delta_{\vec{k}}$ are shown in Figs.~\ref{fig:result}(c) and ~\ref{fig:result}(d) for two different doping concentrations and $\tilde{t}=20~\mathrm{meV}$ and the other parameters taken from the first-principles calculations and the RMFT solutions. The Fermi surfaces and the nodal lines avoid intersecting at low doping concentrations and with a relatively large $\tilde{t}$, therefore the system has a full SC gap. The density of states is shown in Fig.~\ref{fig:result}(e). Otherwise, the quasiparticle excitations are gapless at the intersection points of the Fermi surfaces and the nodal lines of the gap function, which is shown in Fig.~\ref{fig:result}(f). The phase diagram in the $\tilde{t}$-doping $\delta$ plane is shown in Fig.~\ref{fig:result}(b). Therefore, the observed full and nodal SC gaps in different spatial regions can be explained by the inhomogeneous doping concentrations.

The spin-dependent hopping $\tilde{t}$ changes the band structure and reduces the DOS on the Fermi surface, thus one may expect that the pairing order parameter $\Delta$ (and also $T_c$) will be strongly suppressed. This is not true if $\tilde{t}$ is relatively weak compared with $J$. The electron pairing is driven by the superexchange interaction $J$, so the pairing order parameter and $T_c$ are controlled by the DOS around the Fermi level within the energy range of order $J$. Even though a relatively weak spin-dependent hopping $\tilde{t}$ (compared to $J$) will redistribute the DOS within the energy range of $\tilde{t}$, the total DOS within the range of $J$ will not be changed dramatically. Therefore, the pairing order parameter and $T_c$ are not significantly changed. This is confirmed by our RMFT calculations in the presence of $\tilde{t}$. Taking a doping concentration 0.05 as an example, we find that $\Delta= 0.148$ in the absence of $\tilde{t}$, and $\Delta= 0.138$ with $\tilde{t}= 50$ meV.

\section{Summary and Discussion}

To summarize, we show that for a copper-oxide monolayer grown on a Bi2212 substrate, the T-CuO/Bi2212 structure is more stable than the previously postulated CuO$_{2}$/Bi2212 structure. The T-CuO monolayer consists of two CuO$_{2}$ layers sharing the same O atoms. We argue that one of the Cu sublattice remains half filled and may be AF ordered, while the other sublattice is hole doped and superconducting. The proximity of the SC sublattice to the AF sublattice can give rise to a full SC gap, which provides an explanation for the experiments by Zhong {\it et al.} \cite{zhong2016nodeless}, even though the SC pairing has a $d$-wave symmetry.

In our scenario, a full SC gap is induced from the proximity to the AF ordered half filled CuO$_{2}$ layer. One physical consequence of this scenario is the band folding induced by spin-dependent hopping from the AF order, which might be observed by ARPES. Even if the AF order is not truly long range and static at finite temperatures due to the Mermin-Wagner theorem \cite{mermin1966absence}, its correlation length $\xi_{\mathrm{AF}}$ diverges exponentially approaching zero temperature in the renormalized classical regime \cite{chakravarty1988low, chakravarty1989two}. As long as $\xi_{\mathrm{AF}}$ is larger than the SC coherence length, we expect that the spin-dependent hopping within $\xi_{AF}$ can still induce a full SC gap.

\acknowledgements
We thank Dung-Hai Lee for enlightening discussions. The numerical simulations were performed on Tianhe-I Supercomputer System in Tianjin and on Tianhe-II Supercomputer System in Guangzhou. This work was supported by the National Natural Science Foundation of China (Grant No. 11374018) and the National Key Basic Research Program of China (Grant No. 2014CB920902).

\appendix

\section{Details of spin-polarized calculations}\label{A}

In order to determine the exchange parameters $J_d$ and $J$ in the magnetic interaction model (\ref{Hs}), we employ the GGA$+U$ method introduced by Dudarev {\it et al.} \cite{dudarev1998electron} to calculate the ground-state energies of several magnetically ordered states. We set $U=7.5$ eV and $J^\prime=0.98$ eV on the Cu atoms \cite{anisimov1991band}, which corresponds to $U_{\mathrm{eff}}=U-J^\prime=6.52$ eV in Dudarev's approach. The ferromagnetic (FM) order, the N\'eel order, and the stripe AF order are adopted in the calculations, which are shown in the second row of Table~\ref{tab:configurations}. The classical energy per supercell indicated by the rectangles of each configuration from the classical Heisenberg model and the DFT calculations are listed in the third and fourth rows of Table~\ref{tab:configurations}, respectively. Assuming a local moment size $S=\frac{1}{2}$ and fitting the classical energies to the DFT results gives $J_d=-5.6$ meV and $J=119.6$ meV. We also calculate the ground-state energy of the spiral magnetic order, in which the nearest-neighbor Cu spins align vertically and the next-nearest-neighbor Cu spins align anti-parallel to each other. The energy difference between the spiral order and the stripe order is less than 1 meV per atom. This is consistent with the fact that their classical energy from the Heisenberg model (\ref{Hs}) is the same and thus confirms the validity of the Heisenberg model.

These magnetic interaction strengths are close to those in the T-CuO film grown epitaxially on a SrTiO$_{3}$ substrate \cite{moser2014angle, adolphs2016non}, except that the intersublattice interaction $J_{d}$ is weakly FM, which can be explained by the Goodenough-Kanamori rule \cite{goodenough1955theory}.

\begin{table}[h]
\centering
\caption{The local moment configuration and the ground-state energy per supercell for each magnetically ordered state adopted in the spin-polarized DFT$+U$ calculations.}
\label{tab:configurations}
\begin{tabular}{cccc}
\hline\hline
Order & FM & N\'eel & Stripe \\
\hline
Configuration & \includegraphics[width=0.08\textwidth]{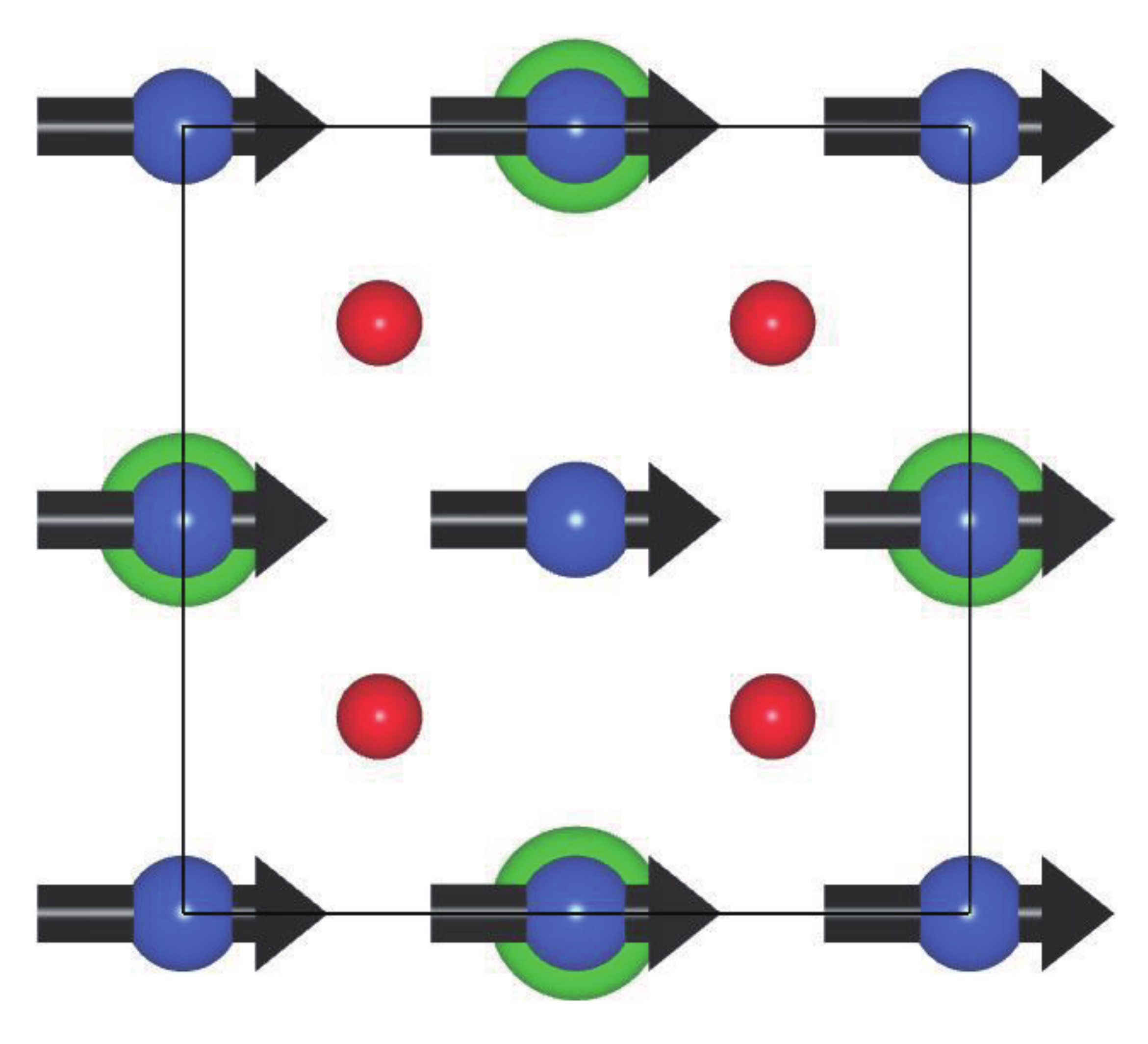} & \includegraphics[width=0.08\textwidth]{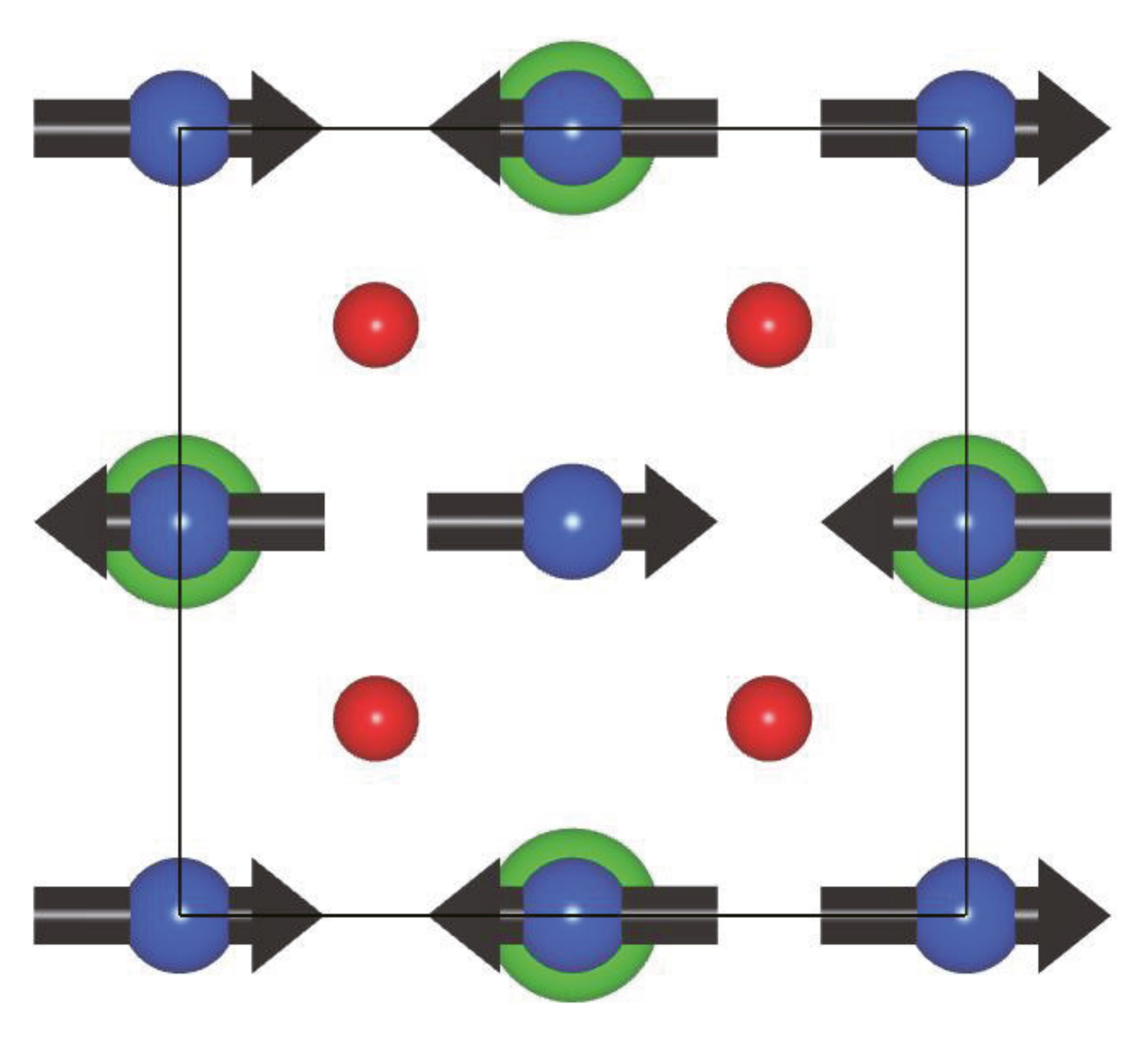} &  \includegraphics[width=0.08\textwidth]{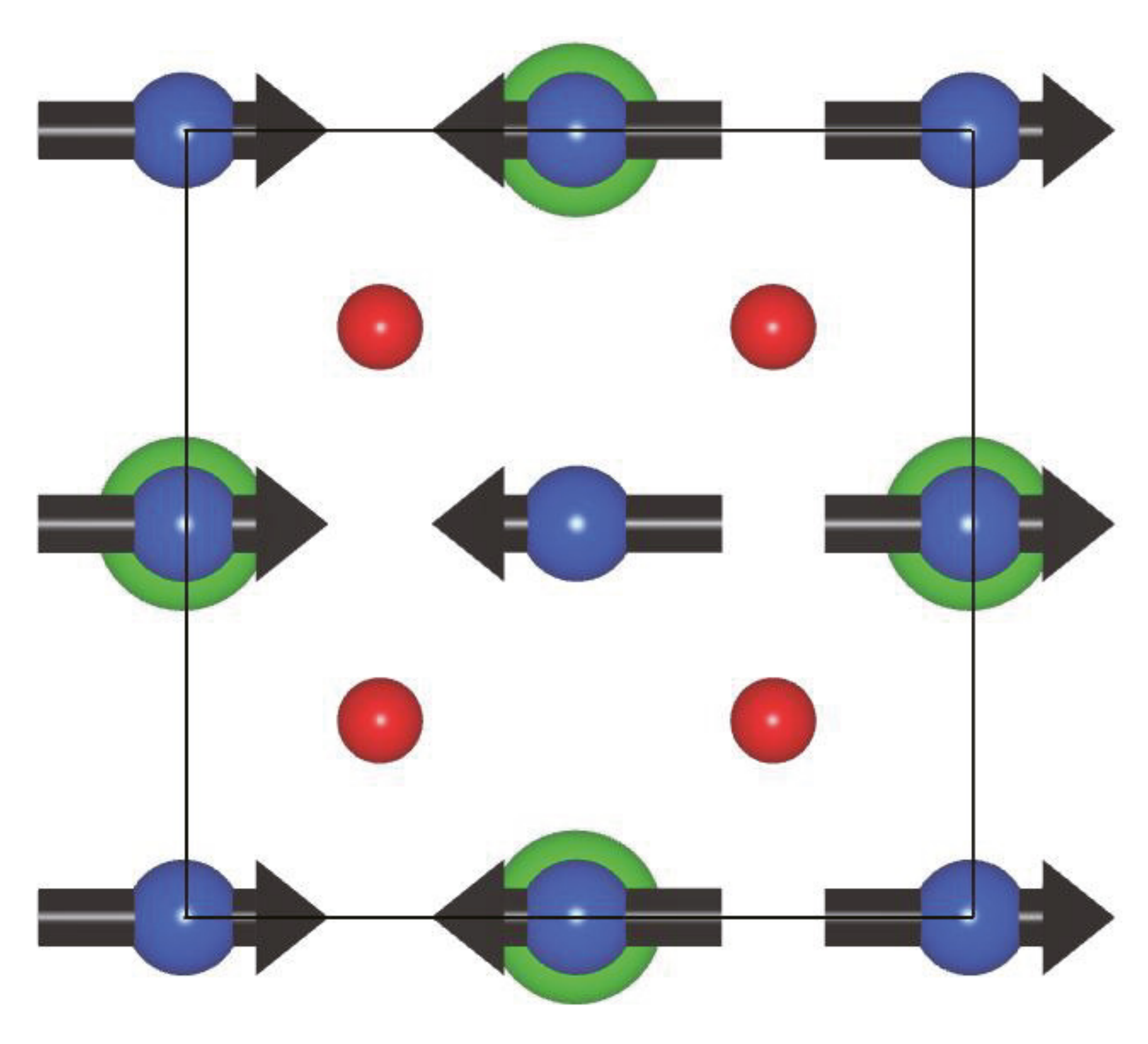} \\
Classical & $c+8(J+J_{d})S^2$& $c+8(J-J_d)S^2$ & $c-8JS^2$ \\
DFT$+U$ (eV) & -346.9635 & -346.9418 & -347.4311 \\
\hline\hline
\end{tabular}
\end{table}

\bibliography{library}

\begin{thebibliography}{38}
\expandafter\ifx\csname natexlab\endcsname\relax\def\natexlab#1{#1}\fi
\expandafter\ifx\csname bibnamefont\endcsname\relax
  \def\bibnamefont#1{#1}\fi
\expandafter\ifx\csname bibfnamefont\endcsname\relax
  \def\bibfnamefont#1{#1}\fi
\expandafter\ifx\csname citenamefont\endcsname\relax
  \def\citenamefont#1{#1}\fi
\expandafter\ifx\csname url\endcsname\relax
  \def\url#1{\texttt{#1}}\fi
\expandafter\ifx\csname urlprefix\endcsname\relax\def\urlprefix{URL }\fi
\providecommand{\bibinfo}[2]{#2}
\providecommand{\eprint}[2][]{\url{#2}}

\bibitem[{\citenamefont{Bednorz and M{\"{u}}ller}(1986)}]{bednorz1986possible}
\bibinfo{author}{\bibfnamefont{J.~G.} \bibnamefont{Bednorz}} \bibnamefont{and}
  \bibinfo{author}{\bibfnamefont{K.~A.} \bibnamefont{M{\"{u}}ller}},
  \bibinfo{journal}{Z. Phys. B: Condens. Matter} \textbf{\bibinfo{volume}{64}},
  \bibinfo{pages}{189} (\bibinfo{year}{1986}).

\bibitem[{\citenamefont{Timusk and Statt}(1999)}]{timusk1999pseudogap}
\bibinfo{author}{\bibfnamefont{T.}~\bibnamefont{Timusk}} \bibnamefont{and}
  \bibinfo{author}{\bibfnamefont{B.}~\bibnamefont{Statt}},
  \bibinfo{journal}{Rep. Prog. Phys.} \textbf{\bibinfo{volume}{62}},
  \bibinfo{pages}{61} (\bibinfo{year}{1999}).

\bibitem[{\citenamefont{Keimer et~al.}(2015)\citenamefont{Keimer, Kivelson,
  Norman, Uchida, and Zaanen}}]{keimer2015quantum}
\bibinfo{author}{\bibfnamefont{B.}~\bibnamefont{Keimer}},
  \bibinfo{author}{\bibfnamefont{S.~A.} \bibnamefont{Kivelson}},
  \bibinfo{author}{\bibfnamefont{M.~R.} \bibnamefont{Norman}},
  \bibinfo{author}{\bibfnamefont{S.}~\bibnamefont{Uchida}}, \bibnamefont{and}
  \bibinfo{author}{\bibfnamefont{J.}~\bibnamefont{Zaanen}},
  \bibinfo{journal}{Nature (London)} \textbf{\bibinfo{volume}{518}},
  \bibinfo{pages}{179} (\bibinfo{year}{2015}).

\bibitem[{\citenamefont{Zaanen et~al.}(2006)\citenamefont{Zaanen, Chakravarty,
  Senthil, Anderson, Lee, Schmalian, Imada, Pines, Randeria, Varma
  et~al.}}]{zaanen2006towards}
\bibinfo{author}{\bibfnamefont{J.}~\bibnamefont{Zaanen}},
  \bibinfo{author}{\bibfnamefont{S.}~\bibnamefont{Chakravarty}},
  \bibinfo{author}{\bibfnamefont{T.}~\bibnamefont{Senthil}},
  \bibinfo{author}{\bibfnamefont{P.}~\bibnamefont{Anderson}},
  \bibinfo{author}{\bibfnamefont{P.}~\bibnamefont{Lee}},
  \bibinfo{author}{\bibfnamefont{J.}~\bibnamefont{Schmalian}},
  \bibinfo{author}{\bibfnamefont{M.}~\bibnamefont{Imada}},
  \bibinfo{author}{\bibfnamefont{D.}~\bibnamefont{Pines}},
  \bibinfo{author}{\bibfnamefont{M.}~\bibnamefont{Randeria}},
  \bibinfo{author}{\bibfnamefont{C.~M.} \bibnamefont{Varma}},
  \bibnamefont{et~al.}, \bibinfo{journal}{Nat. Phys.}
  \textbf{\bibinfo{volume}{2}}, \bibinfo{pages}{138} (\bibinfo{year}{2006}).

\bibitem[{\citenamefont{Damascelli et~al.}(2003)\citenamefont{Damascelli,
  Hussain, and Shen}}]{damascelli2003angle}
\bibinfo{author}{\bibfnamefont{A.}~\bibnamefont{Damascelli}},
  \bibinfo{author}{\bibfnamefont{Z.}~\bibnamefont{Hussain}}, \bibnamefont{and}
  \bibinfo{author}{\bibfnamefont{Z.-X.} \bibnamefont{Shen}},
  \bibinfo{journal}{Rev. Mod. Phys.} \textbf{\bibinfo{volume}{75}},
  \bibinfo{pages}{473} (\bibinfo{year}{2003}).

\bibitem[{\citenamefont{Fischer et~al.}(2007)\citenamefont{Fischer, Kugler,
  Maggio-Aprile, Berthod, and Renner}}]{fischer2007scanning}
\bibinfo{author}{\bibfnamefont{{\O}.}~\bibnamefont{Fischer}},
  \bibinfo{author}{\bibfnamefont{M.}~\bibnamefont{Kugler}},
  \bibinfo{author}{\bibfnamefont{I.}~\bibnamefont{Maggio-Aprile}},
  \bibinfo{author}{\bibfnamefont{C.}~\bibnamefont{Berthod}}, \bibnamefont{and}
  \bibinfo{author}{\bibfnamefont{C.}~\bibnamefont{Renner}},
  \bibinfo{journal}{Rev. Mod. Phys.} \textbf{\bibinfo{volume}{79}},
  \bibinfo{pages}{353} (\bibinfo{year}{2007}).

\bibitem[{\citenamefont{Tsuei and Kirtley}(2000)}]{tsuei2000pairing}
\bibinfo{author}{\bibfnamefont{C.~C.} \bibnamefont{Tsuei}} \bibnamefont{and}
  \bibinfo{author}{\bibfnamefont{J.~R.} \bibnamefont{Kirtley}},
  \bibinfo{journal}{Rev. Mod. Phys.} \textbf{\bibinfo{volume}{72}},
  \bibinfo{pages}{969} (\bibinfo{year}{2000}).

\bibitem[{\citenamefont{Kotliar and Liu}(1988)}]{kotliar1988superexchange}
\bibinfo{author}{\bibfnamefont{G.}~\bibnamefont{Kotliar}} \bibnamefont{and}
  \bibinfo{author}{\bibfnamefont{J.}~\bibnamefont{Liu}},
  \bibinfo{journal}{Phys. Rev. B} \textbf{\bibinfo{volume}{38}},
  \bibinfo{pages}{5142} (\bibinfo{year}{1988}).

\bibitem[{\citenamefont{Zhang et~al.}(1988)\citenamefont{Zhang, Gros, Rice, and
  Shiba}}]{zhang1988renormalised}
\bibinfo{author}{\bibfnamefont{F.~C.} \bibnamefont{Zhang}},
  \bibinfo{author}{\bibfnamefont{C.}~\bibnamefont{Gros}},
  \bibinfo{author}{\bibfnamefont{T.~M.} \bibnamefont{Rice}}, \bibnamefont{and}
  \bibinfo{author}{\bibfnamefont{H.}~\bibnamefont{Shiba}},
  \bibinfo{journal}{Supercond. Sci. Technol.} \textbf{\bibinfo{volume}{1}},
  \bibinfo{pages}{36} (\bibinfo{year}{1988}).

\bibitem[{\citenamefont{Scalapino et~al.}(1986)\citenamefont{Scalapino, {Loh,
  Jr.}, and Hirsch}}]{scalapino1986d}
\bibinfo{author}{\bibfnamefont{D.~J.} \bibnamefont{Scalapino}},
  \bibinfo{author}{\bibfnamefont{E.}~\bibnamefont{{Loh, Jr.}}},
  \bibnamefont{and} \bibinfo{author}{\bibfnamefont{J.~E.}
  \bibnamefont{Hirsch}}, \bibinfo{journal}{Phys. Rev. B}
  \textbf{\bibinfo{volume}{34}}, \bibinfo{pages}{8190} (\bibinfo{year}{1986}).

\bibitem[{\citenamefont{Zhong et~al.}(2016)\citenamefont{Zhong, Wang, Han, Lv,
  Wang, Zhang, Ding, Zhang, Wang, He et~al.}}]{zhong2016nodeless}
\bibinfo{author}{\bibfnamefont{Y.}~\bibnamefont{Zhong}},
  \bibinfo{author}{\bibfnamefont{Y.}~\bibnamefont{Wang}},
  \bibinfo{author}{\bibfnamefont{S.}~\bibnamefont{Han}},
  \bibinfo{author}{\bibfnamefont{Y.-F.} \bibnamefont{Lv}},
  \bibinfo{author}{\bibfnamefont{W.-L.} \bibnamefont{Wang}},
  \bibinfo{author}{\bibfnamefont{D.}~\bibnamefont{Zhang}},
  \bibinfo{author}{\bibfnamefont{H.}~\bibnamefont{Ding}},
  \bibinfo{author}{\bibfnamefont{Y.-M.} \bibnamefont{Zhang}},
  \bibinfo{author}{\bibfnamefont{L.}~\bibnamefont{Wang}},
  \bibinfo{author}{\bibfnamefont{K.}~\bibnamefont{He}}, \bibnamefont{et~al.},
  \bibinfo{journal}{Sci. Bull.} \textbf{\bibinfo{volume}{61}},
  \bibinfo{pages}{1239} (\bibinfo{year}{2016}).

\bibitem[{\citenamefont{Zhu et~al.}(2016)\citenamefont{Zhu, Zhang, and
  Zhang}}]{zhu2016proximity}
\bibinfo{author}{\bibfnamefont{G.-Y.} \bibnamefont{Zhu}},
  \bibinfo{author}{\bibfnamefont{F.-C.} \bibnamefont{Zhang}}, \bibnamefont{and}
  \bibinfo{author}{\bibfnamefont{G.-M.} \bibnamefont{Zhang}},
  \bibinfo{journal}{Phys. Rev. B} \textbf{\bibinfo{volume}{94}},
  \bibinfo{pages}{174501} (\bibinfo{year}{2016}).

\bibitem[{\citenamefont{Wang et~al.}(2017)\citenamefont{Wang, Wang, and
  Chen}}]{wang2017nodal}
\bibinfo{author}{\bibfnamefont{Y.}~\bibnamefont{Wang}},
  \bibinfo{author}{\bibfnamefont{Z.-H.} \bibnamefont{Wang}}, \bibnamefont{and}
  \bibinfo{author}{\bibfnamefont{W.-Q.} \bibnamefont{Chen}},
  \bibinfo{journal}{Physical Review B} \textbf{\bibinfo{volume}{96}},
  \bibinfo{pages}{104507} (\bibinfo{year}{2017}).

\bibitem[{\citenamefont{Zhu et~al.}(2017)\citenamefont{Zhu, Wang, and
  Zhang}}]{zhu2017two}
\bibinfo{author}{\bibfnamefont{G.-Y.} \bibnamefont{Zhu}},
  \bibinfo{author}{\bibfnamefont{Z.}~\bibnamefont{Wang}}, \bibnamefont{and}
  \bibinfo{author}{\bibfnamefont{G.-M.} \bibnamefont{Zhang}},
  \bibinfo{journal}{Europhys. Lett.} \textbf{\bibinfo{volume}{118}},
  \bibinfo{pages}{37004} (\bibinfo{year}{2017}).

\bibitem[{\citenamefont{Zhu and Zhang}(2017)}]{zhu2017topological}
\bibinfo{author}{\bibfnamefont{G.-Y.} \bibnamefont{Zhu}} \bibnamefont{and}
  \bibinfo{author}{\bibfnamefont{G.-M.} \bibnamefont{Zhang}},
  \bibinfo{journal}{Europhys. Lett.} \textbf{\bibinfo{volume}{117}},
  \bibinfo{pages}{67007} (\bibinfo{year}{2017}).

\bibitem[{\citenamefont{Lu et~al.}(2014)\citenamefont{Lu, Xiang, and
  Lee}}]{Lu2014}
\bibinfo{author}{\bibfnamefont{Y.-M.} \bibnamefont{Lu}},
  \bibinfo{author}{\bibfnamefont{T.}~\bibnamefont{Xiang}}, \bibnamefont{and}
  \bibinfo{author}{\bibfnamefont{D.-H.} \bibnamefont{Lee}},
  \bibinfo{journal}{Nat. Phys.} \textbf{\bibinfo{volume}{10}},
  \bibinfo{pages}{634} (\bibinfo{year}{2014}).

\bibitem[{\citenamefont{Bl{\"{o}}chl}(1994)}]{blochl1994projector}
\bibinfo{author}{\bibfnamefont{P.~E.} \bibnamefont{Bl{\"{o}}chl}},
  \bibinfo{journal}{Phys. Rev. B} \textbf{\bibinfo{volume}{50}},
  \bibinfo{pages}{17953} (\bibinfo{year}{1994}).

\bibitem[{\citenamefont{Kresse and Joubert}(1999)}]{kresse1999from}
\bibinfo{author}{\bibfnamefont{G.}~\bibnamefont{Kresse}} \bibnamefont{and}
  \bibinfo{author}{\bibfnamefont{D.}~\bibnamefont{Joubert}},
  \bibinfo{journal}{Phys. Rev. B} \textbf{\bibinfo{volume}{59}},
  \bibinfo{pages}{1758} (\bibinfo{year}{1999}).

\bibitem[{\citenamefont{Kresse and
  Furthm{\"{u}}ller}(1996{\natexlab{a}})}]{kresse1996efficiency}
\bibinfo{author}{\bibfnamefont{G.}~\bibnamefont{Kresse}} \bibnamefont{and}
  \bibinfo{author}{\bibfnamefont{J.}~\bibnamefont{Furthm{\"{u}}ller}},
  \bibinfo{journal}{Comput. Mater. Sci.} \textbf{\bibinfo{volume}{6}},
  \bibinfo{pages}{15} (\bibinfo{year}{1996}{\natexlab{a}}).

\bibitem[{\citenamefont{Kresse and
  Furthm{\"{u}}ller}(1996{\natexlab{b}})}]{kresse1996efficient}
\bibinfo{author}{\bibfnamefont{G.}~\bibnamefont{Kresse}} \bibnamefont{and}
  \bibinfo{author}{\bibfnamefont{J.}~\bibnamefont{Furthm{\"{u}}ller}},
  \bibinfo{journal}{Phys. Rev. B} \textbf{\bibinfo{volume}{54}},
  \bibinfo{pages}{11169} (\bibinfo{year}{1996}{\natexlab{b}}).

\bibitem[{\citenamefont{Perdew et~al.}(1996)\citenamefont{Perdew, Burke, and
  Ernzerhof}}]{perdew1996generalized}
\bibinfo{author}{\bibfnamefont{J.~P.} \bibnamefont{Perdew}},
  \bibinfo{author}{\bibfnamefont{K.}~\bibnamefont{Burke}}, \bibnamefont{and}
  \bibinfo{author}{\bibfnamefont{M.}~\bibnamefont{Ernzerhof}},
  \bibinfo{journal}{Phys. Rev. Lett.} \textbf{\bibinfo{volume}{77}},
  \bibinfo{pages}{3865} (\bibinfo{year}{1996}).

\bibitem[{\citenamefont{Sunshine et~al.}(1988)\citenamefont{Sunshine, Siegrist,
  Schneemeyer, Murphy, Cava, Batlogg, {Van Dover}, Fleming, Glarum, Nakahara
  et~al.}}]{sunshine1988structure}
\bibinfo{author}{\bibfnamefont{S.~A.} \bibnamefont{Sunshine}},
  \bibinfo{author}{\bibfnamefont{T.}~\bibnamefont{Siegrist}},
  \bibinfo{author}{\bibfnamefont{L.~F.} \bibnamefont{Schneemeyer}},
  \bibinfo{author}{\bibfnamefont{D.~W.} \bibnamefont{Murphy}},
  \bibinfo{author}{\bibfnamefont{R.~J.} \bibnamefont{Cava}},
  \bibinfo{author}{\bibfnamefont{B.}~\bibnamefont{Batlogg}},
  \bibinfo{author}{\bibfnamefont{R.~B.} \bibnamefont{{Van Dover}}},
  \bibinfo{author}{\bibfnamefont{R.~M.} \bibnamefont{Fleming}},
  \bibinfo{author}{\bibfnamefont{S.~H.} \bibnamefont{Glarum}},
  \bibinfo{author}{\bibfnamefont{S.}~\bibnamefont{Nakahara}},
  \bibnamefont{et~al.}, \bibinfo{journal}{Phys. Rev. B}
  \textbf{\bibinfo{volume}{38}}, \bibinfo{pages}{893} (\bibinfo{year}{1988}).

\bibitem[{\citenamefont{Monkhorst and Pack}(1976)}]{monkhorst1976special}
\bibinfo{author}{\bibfnamefont{H.~J.} \bibnamefont{Monkhorst}}
  \bibnamefont{and} \bibinfo{author}{\bibfnamefont{J.~D.} \bibnamefont{Pack}},
  \bibinfo{journal}{Phys. Rev. B} \textbf{\bibinfo{volume}{13}},
  \bibinfo{pages}{5188} (\bibinfo{year}{1976}).

\bibitem[{\citenamefont{Zhang and Rice}(1988)}]{zhang1988effective}
\bibinfo{author}{\bibfnamefont{F.~C.} \bibnamefont{Zhang}} \bibnamefont{and}
  \bibinfo{author}{\bibfnamefont{T.~M.} \bibnamefont{Rice}},
  \bibinfo{journal}{Phys. Rev. B} \textbf{\bibinfo{volume}{37}},
  \bibinfo{pages}{3759} (\bibinfo{year}{1988}).

\bibitem[{\citenamefont{Andersen et~al.}(1995)\citenamefont{Andersen,
  Liechtenstein, Jepsen, and Paulsen}}]{andersen1995lda}
\bibinfo{author}{\bibfnamefont{O.~K.} \bibnamefont{Andersen}},
  \bibinfo{author}{\bibfnamefont{A.~I.} \bibnamefont{Liechtenstein}},
  \bibinfo{author}{\bibfnamefont{O.}~\bibnamefont{Jepsen}}, \bibnamefont{and}
  \bibinfo{author}{\bibfnamefont{F.}~\bibnamefont{Paulsen}},
  \bibinfo{journal}{J. Phys. Chem. Solids} \textbf{\bibinfo{volume}{56}},
  \bibinfo{pages}{1573} (\bibinfo{year}{1995}).

\bibitem[{\citenamefont{Moser et~al.}(2014)\citenamefont{Moser, Moreschini,
  Yang, Innocenti, Fuchs, Hansen, Chang, Kim, Walter, Bostwick
  et~al.}}]{moser2014angle}
\bibinfo{author}{\bibfnamefont{S.}~\bibnamefont{Moser}},
  \bibinfo{author}{\bibfnamefont{L.}~\bibnamefont{Moreschini}},
  \bibinfo{author}{\bibfnamefont{H.-Y.} \bibnamefont{Yang}},
  \bibinfo{author}{\bibfnamefont{D.}~\bibnamefont{Innocenti}},
  \bibinfo{author}{\bibfnamefont{F.}~\bibnamefont{Fuchs}},
  \bibinfo{author}{\bibfnamefont{N.~H.} \bibnamefont{Hansen}},
  \bibinfo{author}{\bibfnamefont{Y.~J.} \bibnamefont{Chang}},
  \bibinfo{author}{\bibfnamefont{K.~S.} \bibnamefont{Kim}},
  \bibinfo{author}{\bibfnamefont{A.~L.} \bibnamefont{Walter}},
  \bibinfo{author}{\bibfnamefont{A.}~\bibnamefont{Bostwick}},
  \bibnamefont{et~al.}, \bibinfo{journal}{Phys. Rev. Lett.}
  \textbf{\bibinfo{volume}{113}}, \bibinfo{pages}{187001}
  (\bibinfo{year}{2014}).

\bibitem[{\citenamefont{Adolphs et~al.}(2016)\citenamefont{Adolphs, Moser,
  Sawatzky, and Berciu}}]{adolphs2016non}
\bibinfo{author}{\bibfnamefont{C.~P.~J.} \bibnamefont{Adolphs}},
  \bibinfo{author}{\bibfnamefont{S.}~\bibnamefont{Moser}},
  \bibinfo{author}{\bibfnamefont{G.~A.} \bibnamefont{Sawatzky}},
  \bibnamefont{and} \bibinfo{author}{\bibfnamefont{M.}~\bibnamefont{Berciu}},
  \bibinfo{journal}{Phys. Rev. Lett.} \textbf{\bibinfo{volume}{116}},
  \bibinfo{pages}{087002} (\bibinfo{year}{2016}).

\bibitem[{\citenamefont{Gutzwiller}(1963)}]{gutzwiller1963effect}
\bibinfo{author}{\bibfnamefont{M.~C.} \bibnamefont{Gutzwiller}},
  \bibinfo{journal}{Phys. Rev. Lett.} \textbf{\bibinfo{volume}{10}},
  \bibinfo{pages}{159} (\bibinfo{year}{1963}).

\bibitem[{\citenamefont{Anderson}(1987)}]{anderson1987resonating}
\bibinfo{author}{\bibfnamefont{P.~W.} \bibnamefont{Anderson}},
  \bibinfo{journal}{Science} \textbf{\bibinfo{volume}{235}},
  \bibinfo{pages}{1196} (\bibinfo{year}{1987}).

\bibitem[{\citenamefont{Brinkmann and Rice}(1970)}]{brinkmann1970application}
\bibinfo{author}{\bibfnamefont{W.}~\bibnamefont{Brinkmann}} \bibnamefont{and}
  \bibinfo{author}{\bibfnamefont{T.}~\bibnamefont{Rice}},
  \bibinfo{journal}{Phys. Rev. B} \textbf{\bibinfo{volume}{2}},
  \bibinfo{pages}{4302} (\bibinfo{year}{1970}).

\bibitem[{\citenamefont{Anisimov et~al.}(2002)\citenamefont{Anisimov, Nekrasov,
  Kondakov, Rice, and Sigrist}}]{anisimov2002orbital}
\bibinfo{author}{\bibfnamefont{V.~I.} \bibnamefont{Anisimov}},
  \bibinfo{author}{\bibfnamefont{I.~A.} \bibnamefont{Nekrasov}},
  \bibinfo{author}{\bibfnamefont{D.~E.} \bibnamefont{Kondakov}},
  \bibinfo{author}{\bibfnamefont{T.~M.} \bibnamefont{Rice}}, \bibnamefont{and}
  \bibinfo{author}{\bibfnamefont{M.}~\bibnamefont{Sigrist}},
  \bibinfo{journal}{Eur. Phys. J. B} \textbf{\bibinfo{volume}{25}},
  \bibinfo{pages}{191} (\bibinfo{year}{2002}).

\bibitem[{\citenamefont{J\ifmmode~\mbox{\c{e}}\else \c{e}\fi{}drak and
  Spa\l{}ek}(2010)}]{j?drak2010consistent}
\bibinfo{author}{\bibfnamefont{J.}~\bibnamefont{J\ifmmode~\mbox{\c{e}}\else
  \c{e}\fi{}drak}} \bibnamefont{and}
  \bibinfo{author}{\bibfnamefont{J.}~\bibnamefont{Spa\l{}ek}},
  \bibinfo{journal}{Phys. Rev. B} \textbf{\bibinfo{volume}{81}},
  \bibinfo{pages}{073108} (\bibinfo{year}{2010}).

\bibitem[{\citenamefont{Mermin and Wagner}(1966)}]{mermin1966absence}
\bibinfo{author}{\bibfnamefont{N.~D.} \bibnamefont{Mermin}} \bibnamefont{and}
  \bibinfo{author}{\bibfnamefont{H.}~\bibnamefont{Wagner}},
  \bibinfo{journal}{Phys. Rev. Lett.} \textbf{\bibinfo{volume}{17}},
  \bibinfo{pages}{1133} (\bibinfo{year}{1966}).

\bibitem[{\citenamefont{Chakravarty et~al.}(1988)\citenamefont{Chakravarty,
  Halperin, and Nelson}}]{chakravarty1988low}
\bibinfo{author}{\bibfnamefont{S.}~\bibnamefont{Chakravarty}},
  \bibinfo{author}{\bibfnamefont{B.~I.} \bibnamefont{Halperin}},
  \bibnamefont{and} \bibinfo{author}{\bibfnamefont{D.~R.}
  \bibnamefont{Nelson}}, \bibinfo{journal}{Phys. Rev. Lett.}
  \textbf{\bibinfo{volume}{60}}, \bibinfo{pages}{1057} (\bibinfo{year}{1988}).

\bibitem[{\citenamefont{Chakravarty et~al.}(1989)\citenamefont{Chakravarty,
  Halperin, and Nelson}}]{chakravarty1989two}
\bibinfo{author}{\bibfnamefont{S.}~\bibnamefont{Chakravarty}},
  \bibinfo{author}{\bibfnamefont{B.~I.} \bibnamefont{Halperin}},
  \bibnamefont{and} \bibinfo{author}{\bibfnamefont{D.~R.}
  \bibnamefont{Nelson}}, \bibinfo{journal}{Phys. Rev. B}
  \textbf{\bibinfo{volume}{39}}, \bibinfo{pages}{2344} (\bibinfo{year}{1989}).

\bibitem[{\citenamefont{Dudarev et~al.}(1998)\citenamefont{Dudarev, Botton,
  Savrasov, Humphreys, and Sutton}}]{dudarev1998electron}
\bibinfo{author}{\bibfnamefont{S.~L.} \bibnamefont{Dudarev}},
  \bibinfo{author}{\bibfnamefont{G.~A.} \bibnamefont{Botton}},
  \bibinfo{author}{\bibfnamefont{S.~Y.} \bibnamefont{Savrasov}},
  \bibinfo{author}{\bibfnamefont{C.~J.} \bibnamefont{Humphreys}},
  \bibnamefont{and} \bibinfo{author}{\bibfnamefont{A.~P.}
  \bibnamefont{Sutton}}, \bibinfo{journal}{Phys. Rev. B}
  \textbf{\bibinfo{volume}{57}}, \bibinfo{pages}{1505} (\bibinfo{year}{1998}).

\bibitem[{\citenamefont{Anisimov et~al.}(1991)\citenamefont{Anisimov, Zaanen,
  and Andersen}}]{anisimov1991band}
\bibinfo{author}{\bibfnamefont{V.~I.} \bibnamefont{Anisimov}},
  \bibinfo{author}{\bibfnamefont{J.}~\bibnamefont{Zaanen}}, \bibnamefont{and}
  \bibinfo{author}{\bibfnamefont{O.~K.} \bibnamefont{Andersen}},
  \bibinfo{journal}{Phys. Rev. B} \textbf{\bibinfo{volume}{44}},
  \bibinfo{pages}{943} (\bibinfo{year}{1991}).

\bibitem[{\citenamefont{Goodenough}(1955)}]{goodenough1955theory}
\bibinfo{author}{\bibfnamefont{J.~B.} \bibnamefont{Goodenough}},
  \bibinfo{journal}{Phys. Rev.} \textbf{\bibinfo{volume}{100}},
  \bibinfo{pages}{564} (\bibinfo{year}{1955}).

\end{thebibliography}

\end{document}